\documentclass{elsart4-1}
\input epsf

\usepackage{amssymb}

\usepackage[english,francais]{babel}

\newtheorem{e-proposition}[theorem]{Proposition}

\newtheorem{e-definition}[theorem]{Definition\rm}

\renewcommand{\theequation}{\arabic{equation}}
\def\og{\leavevmode\raise.3ex\hbox{$\scriptscriptstyle\langle\!\langle$~}}
\def\fg{\leavevmode\raise.3ex\hbox{~$\!\scriptscriptstyle\,\rangle\!\rangle$}}

\begin{document}

\def\ra{\rangle}
\def\la{\langle}
\def\be{\begin{equation}}
\def\ee{\end{equation}}
\def\CO{\mathcal{O}}
\def\CM{\mathcal{M}}
\def\CV{\mathcal{V}}
\def\x{\xi}
\def\Tr{\rm Tr}
\def\s{{\sigma}}
\def\rh{{\rho}}
\def\t{{\tau}}
\def\tt{{\tilde{\tau}}}
\def\l{{\lambda}}
\def\D{\Delta}
\def\m{\mu}

\begin{frontmatter}


\selectlanguage{english}
\title{Free Field Theory as a String Theory?}




\selectlanguage{english}
\author{Rajesh Gopakumar}
\ead{gopakumr@mri.ernet.in}

\address{Harish-Chandra Research Institute, Chhatnag Rd., Jhusi, 
Allahabad, India 211019}

\begin{abstract}
An approach to systematically implement open-closed string duality 
for free large $N$ gauge theories is summarised. We show how 
the relevant closed string moduli space emerges from a reorganisation
of the Feynman diagrams contributing to free field correlators. We also 
indicate why the resulting integrand on moduli space 
has the right features to be that of a string theory on $AdS$.

\vskip 0.5\baselineskip


\end{abstract}
\end{frontmatter}


\selectlanguage{english}
\section{Introduction: Two Questions}
The picture of 'tHooft's double line diagrams (or open string diagrams)
getting glued up into closed string worldsheets in the large $N$ limit
seems to be borne out in the few examples of the gauge-string duality
that we concretely understand. Making this picture precise is essential 
if we are to obtain, say, a string dual to realistic gauge theories. 
To this end we will consider a couple of questions:
\begin{itemize}
\item{}
How {\it exactly} does a large $N$ field theory reorganise itself into a 
dual closed string theory?
\item{}
Can we {\it systematically} 
construct the closed string theory starting from the 
field theory?
\end{itemize}
We will address these questions in the simplest of contexts:
that of {\it free} field theories in the large $N$ limit. 
Though we will be always keeping an eye on the extensibility of our results 
to the case of non-zero 'tHooft coupling $\lambda$, at least
in a perturbative expansion. 

The general expectation from gauge-string duality is that 
\be
\la \CO_1(k_1)\ldots \CO_n(k_n)\ra_{g}
=\int_{\CM_{g,n}}\la \CV_1(k_1,\x_1)\ldots \CV_n(k_n,\x_n)\ra_{WS}.
\label{opclos}
\ee 
On the left hand side, the
$\CO_i$ are gauge invariant operators and the subscript $g$ refers to 
the contribution to the correlator from Feynman 
diagrams of genus $g$. Recall that 
the ${1\over N}$ expansion helps us isolate contributions of a given genus. 
On the right hand side 
are the corresponding vertex operators $\CV_i$ of the dual string theory.
The subscript $WS$ refers to the averaging with respect to a worldsheet
sigma model action. There is then a further integration 
over the moduli space $\CM_{g,n}$ of genus $g$ 
Riemann surfaces with $n$ marked punctures (labelled by the $\x_i$).

Can we somehow recast the left hand side into the form we expect on 
the right hand side?
This is, in essence, what our pair of questions amount to.
Addressing the first question, we will see that there is a 
simple way to organise the different Feynman diagram contributions
to the free field $n$-point correlation function
so that the nett sum can be written as an integral over 
the moduli space of an $n$-punctured Riemann surface. 

For simplicity of illustration, we will consider (the genus $g$ 
contribution to) 
correlators of free scalar composites ${\rm Tr}\Phi^{J}(k)$. 
\be
G^{\{J_i\}}_g(k_1, k_2,\ldots k_n)=
\la \prod_{i=1}^n \Tr \Phi^{J_i}(k_i)\ra_g.
\label{scorr}
\ee
We will write the individual Feynman diagram 
contributions to this correlator in Schwinger parametrised form.
By reducing the original graphs to a set of ``skeleton'' graphs,
we will argue that the sum over the inequivalent skeleton graphs 
together with the integral over the Schwinger parameters gives precisely
a cell decomposition of the moduli space $\CM_{g,n}\times R_+^n$. 
This gives a very explicit prescription on how to reorganise 
field theory amplitudes into string theory amplitudes. 

Moreover, this prescription also gives us a handle on our second question. 
We will see that the Schwinger parametrised form will enable us to 
write the field theory amplitude Eq.(\ref{scorr}) as 
\be
G^{\{J_i\}}_g(k_1, k_2,\ldots k_n)
=\int_{\CM_{g,n}\times R_+^n}
[d\s]\rh^{(J_i)}(\s)e^{-\sum_{i,j=1}^nk_i\cdot k_jg_{ij}(\s)}.
\ee
The $\s$ collectively denote the moduli of $\CM_{g,n}\times R_+^n$.
The functions $\rh^{(J_i)}(\s)$ and $g_{ij}(\s)$ can be explicitly 
written down. In this form the integrand on moduli space is very reminiscent
of string theory. In fact, as in the expressions for flat space, the
exponential factor is a universal one for all correlators (not just those
of these scalars). All the dependence on the $J_i$ are in the 
multiplicative prefactor $\rh^{(J_i)}(\s)$ which in turn is independent
of the momenta, for this particular class of correlators. For more 
general correlators, the prefactor will contain a polynomial dependence
on the momenta, again as in flat space. 

Our procedure thus gives a candidate for the world sheet correlator 
of vertex operators of the dual string theory\footnote{Strictly 
speaking we would have to carry out the integral over the additional
$R_+^n$ moduli to obtain an integrand on $\CM_{g,n}$. However, as we
will see later, the string theory expression is also naturally extended 
to an integral over $\CM_{g,n}\times R_+^n$,
via a parametrisation of the $n$ external
legs of the vertex operators.}. 
How can we check this hypothesis given that
we don't yet know how to quantise string theory in the kind of highly curved
$AdS$ backgrounds that would presumably be dual to the free field limit? 
We can, as of now, perform a few modest checks. 
Looking at the two and three point 
functions shows that Eq.(\ref{scorr}) gives the corresponding 
correlators in $AdS$ space in a very natural and 
encouraging way\footnote{Though in this case the moduli 
space is trivial and what we are seeing is the $R_+^n$ factor. See previous
footnote.}. One would like to make consistency checks for the four (and 
higher) point functions as well. One very strong check would be to 
verify that the integrand, in these cases, 
satisfies the various properties that are required of a {\it local}
correlator of vertex operators 
in a two dimensional quantum field theory. In particular, 
such a correlator must satisfy the constraints of the 
{\it worldsheet} Operator Product Expansion (OPE). This,
in turn, is manifested in 
the factorisability of amplitudes in spacetime. In the field theory, this 
property is reflected in the {\it spacetime} OPE. We will briefly
indicate some work in progress which aims to follow this logic through,
and support our identification of the integrand.
 
There is also a more fundamental (but less precise) 
reason suggesting that we take the identification, of the integrand 
with a worldsheet correlator, seriously. As we will see, the logic that 
takes us from the field theory diagrams to the stringy moduli
space, in fact, 
implements the geometry underlying open-closed string duality. 
In a sense, it exhibits concretely how the double line diagrams 
get glued up into a closed worldsheet with the holes closing up. 
Therefore we expect that this procedure should also be telling us 
the integrand on the closed string worldsheet. One would like to 
believe then that we have, in all the various worldsheet correlators,
all the information necessary to  
reconstruct the closed string theory. 
Future work will determine how far we can push ahead with this 
answer to our second question.

This being a summary we have tried not 
to get too much into details. Rather, we have expanded here on 
certain broader points. The details, together with a more  
complete set of 
references to related work and other approaches, may be found
in the original papers \cite{fads2}\cite{fads1}. 

\section{Schwinger Parametrisation of Field Theory Amplitudes}

The Schwinger parametric representation of field theory is a well
studied subject. Essentially, one reexpresses the denominator of
all propagators in a Feynman diagram via the identity (appropriate
for Euclidean space correlators) 
\be
{1\over p^2}=\int_0^{\infty} d\tt \exp{\{-\tt p^2\}}.
\label{sch}
\ee 
We can apply this to the individual Feynman graphs (of genus $g$) 
that contribute to Eq.(\ref{scorr}). We obtain
\be
G^{\{J_i\}}_g(k_1, k_2,\ldots k_n)=\sum_{graphs}\int[d^dp]\int_0^{\infty}[d\tt]
e^{-\tilde{P}(k,p,\tt)}.
\ee
Here $\{p\}$ collectively
denote all the independent internal momenta in the loops of the 
Feynman graph, and similarly, $\{\tt\}$ the Schwinger parameters,
one for each internal edge. Since we have repeatedly used Eq.(\ref{sch})
in arriving at this result, it is clear that the exponent 
$\tilde{P}(k,p,\tt)$ is quadratic in all the momenta 
(external as well as internal). 
  
Having converted all the momentum integrals into Gaussian
integrals, we can carry them out explicitly. It is a little
intricate to keep track of the details of the momentum flow. But
the final expressions for an arbitrary Feynman diagram can be
compactly written in graph theoretic terms. For the case of scalar
fields, the expressions can be looked up in field theory textbooks
such as Itzykson and Zuber. The result (in $d$ dimensions) is
\be
G^{\{J_i\}}_g(k_1, k_2,\ldots k_n)
=\sum_{graphs}\int_0^{\infty}{[d\tt]\over \D(\tt)^{d\over 2}}
\exp[-P(\tt,k)].
\label{ttpar}
\ee
The expressions for $P(\tt,k)$ and $\D(\tt)$ are
\be
\D(\tt)=\sum_{T_1}(\prod^{l}\tt).
\label{meas}
\ee
\be
P(\tt, k)=\D(\tt)^{-1}\sum_{T_2}(\prod^{l+1}\tt)(\sum k)^2.
\label{gauss}
\ee
The sum is over various 1-trees
and 2-trees obtained from the original loop diagram. A 1-tree is
obtained by cutting $l$ lines of a diagram with $l$ loops so as to
make a connected tree. While a 2-tree is obtained by cutting $l+1$
lines of the loop so as to form two disjoint trees. Eq.(\ref{meas})
indicates a sum over the set $T_1$ of all 1-trees, with the
product over the $l$ Schwinger parameters of all the cut lines.
The sum over $T_2$ in Eq.(\ref{gauss}) similarly indicates a sum over the
set of all two trees, where the product is over the $\tt$'s of the
$l+1$ cut lines. And $(\sum k)$ is understood to be the sum over
all those external momenta $k_i$ which flow into (either) one of
the two trees. (Note that because of overall momentum
conservation, it does not matter which set of external momenta one
chooses.)

At this stage we do not seem to have accomplished very much 
of a simplification, since we are left with a large number of integrals over
the Schwinger parameters. In fact, since the total number of 
Wick contractions that contribute to Eq.(\ref{scorr}) is 
${1\over 2}\sum_iJ_i$, there are as many propagators and therefore
Schwinger parameters. If the operators we are considering have large $J_i$,
then the corresponding number of integrals is also large. 
If we are to convert this into something universal for all 
$n$-point functions, we have to look for a simplification in this
representation.  

There is indeed such a simplification: though the integral depends naively 
on a large number of Schwinger parameters, the actual non-trivial dependence 
is only on a certain combination of them. To see this, it is best to view 
the Feynman diagrams as double line diagrams or ``fatgraphs''. Between any two
of the external vertices there can be multiple propagators which are 
homotopically deformable into each other (i.e. without crossing other lines). 
Note that viewing the Feynman diagrams in the double line representation
provides an ordering of edges at each vertex and we can unambiguously
speak of edges which are deformable into each other. This is one of the 
places where the underlying non-abelian structure plays a crucial role. 
We will denote by $m_r$, the number of such legs between 
a fixed pair of vertices (and fixed homotopy class) labelled by $r$.
We can then define an ``effective'' Schwinger parameter $\t_r$ 
for this set of edges by
\be
{1 \over \t_r} =\sum_{\m_r=1}^{m_r}{1\over \tt_{r\m_r}}.
\label{resist}
\ee
The simplification is that it is this effective parameter that really 
enters the Schwinger parametrised expressions. It can be shown
\cite{fads2} that
\be
P(\tt,k)=P_{skel}(\t, k),
\label{skelp}
\ee
\be
\D(\tt)={\prod_{r,\m_r}\tt_{r
\m_r}\over\prod_{r}\t_r} \D_{skel}(\t). 
\label{skeld}
\ee
The right hand side of these equations says that one can rewrite the original
expressions, essentially, as a function of the $\t_r$. Moreover, the 
functions that appear on the RHS are defined in exactly the same graph 
theoretic way as 
in Eqs.(\ref{gauss})(\ref{meas}) except that we replace the 
original graph by its {\it skeleton graph}. The skeleton graph is 
obtained from the original one by gluing all the $m_r$ 
homotopic edges into a single edge labelled by $r$. The process is 
illustrated for a sphere level diagram in Fig. 1. This skeleton graph
is the simpler and more universal graph underlying the original
Feynman graph. 

We should mention that the gluing up of 
the Feynman diagram into a skeleton graph can be intuitively
understood from a correspondence between Feynman graphs and electrical 
networks. (This correspondence
was first pointed out in Bjorken's thesis. See \cite{bd}.) 
The essence is that the momenta play the role of currents and
the Schwinger parameters the role of resistances. 
As is evident from Fig. 1, the gluing up of homotopic edges and replacement
by an effective parameter $\t_r$ in Eq.(\ref{resist})
is nothing but parallel resistors being replaced by 
a single effective resistance. From the point of view
of open-closed string duality, this gives us an intuitive picture
of the gluing up of (some of) the holes in the original Feynman graph 
(or open string diagram). Note that the skeleton graph has the same genus
$g$ as the original graph.

\begin{figure}                                 
\begin{center}                                 
\epsfxsize=4.5in
\epsfysize=2.0in                                 
\epsffile{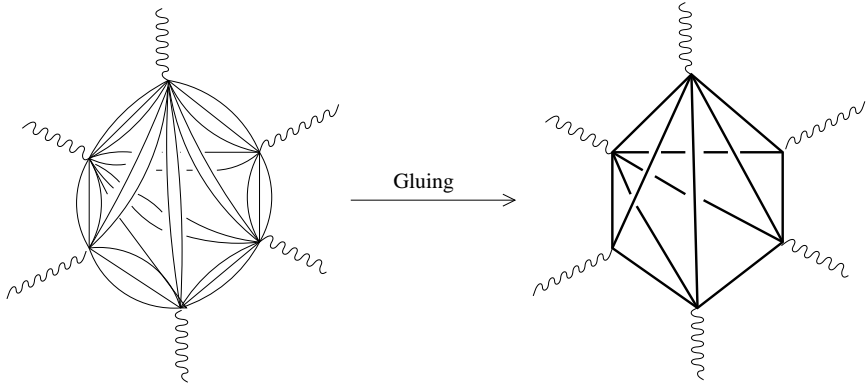}                                 
\end{center}                                 
\caption{Fig.1: Gluing up into a skeleton graph.}
\label{fig:skel}
\end{figure}

Using Eqs. (\ref{skelp}),(\ref{skeld}), for
a given diagram of fixed multiplicity $\{m_r\}$ and connectivity,
we can rewrite the Schwinger integral over the $\tt$'s
as one over the $\t$'s:
\be
\int_0^{\infty}{\prod_{r,\m_r} d\tt_{r\m_r} \over \D(\tt)^{d\over 2}}
e^{-P(\tt,k)} =C^{\{m_r\}}\int_0^{\infty}
\prod_r({d\t_r\over \t_r^{(m_r-1)({d\over2}-1)}}){e^{-P_{skel}(\t, k)}
\over \D_{skel}(\t)^{d\over 2}}.
\label{skel}
\ee
We have also carried out the change of variables in the measure 
from $\tt$ to $\t$ which gives rise to a purely 
 numerical factor $C^{\{m_r\}}$. It depends only on the 
multiplicities and can be explicitly computed \cite{fads2} but will not
be important for us at the moment. Note that the
details of the specific correlator, such as the $J_i$, are contained only 
in the first term (through the dependence on the $m_r$). $P_{skel}$
and $\D_{skel}$ depend only on the topology (connectivity) of the
skeleton graph and are independent of the $J_i$. 

This universality suggests 
that we organise the sum over all graphs in Eq.(\ref{ttpar}) into
a sum over graphs 
having the same underlying skeleton graph 
(but different multiplicities $\{m_r\}$)
and then a sum over various
inequivalent skeleton graphs. The first sum can be carried out
explicitly, since the only dependence on $m_r$ appears in simple prefactors.
We can write the result in the schematic form (the complete answer can 
be found in \cite{fads2})
\be
G^{\{J_i\}}_g(k_1, k_2,\ldots k_n)=\sum_{skel. graphs}
\int_0^{\infty}
{\prod_r d\t_r f^{\{J_i\}}(\t)\over\D_{skel}(\t)^{d\over 2}}
e^{-P_{skel}(\t, k)}.
\label{schem}
\ee
Essentially, the $f^{\{J_i\}}(\t)$ come from carrying out the sum over the
multipicities $m_r$ that are compatible with the same skeleton graph and 
the nett number of fields ${\{J_i\}}$. The sum in Eq.(\ref{schem})
is then over the various inequivalent (i.e. with inequivalent connectivity)
skeleton graphs of genus $g$. 

We have now accomplished the kind
of simplification 
in the Schwinger representation that we were aiming for. 
The number of Schwinger parameters are only as many as the number of edges
of the skeleton graph, a number determined only by $n$ 
and $g$ (and not the $J_i$), as we will soon see. 
By partially gluing up all the homotopic edges 
we have reorganised the Feynman contributions in a way which is more 
universal than the original diagrams. Our $n$-point function has been 
expressed as a sum of contributions from 
the {\it moduli space of skeleton graphs}. 
By which we mean that we are integrating over the lengths $\t$ of the edges 
of the skeleton graphs as well as summing over inequivalent
skeleton graphs. 

\section{From Skeleton Graphs to String Moduli}

How can we characterise this space of skeleton graphs? Like the original 
graph it has $n$ vertices and genus $g$. But having glued together
homotopic edges, the faces of the graphs are at least triangular. In fact, 
for $J_i$ greater than a minimum value (set by $n$), the generic face
is triangular. This is because we can always add extra edges to 
quadrilateral or higher faces and make all faces triangular
without altering the genus $g$. Therefore, 
when we have all possible wick contractions, compatible with
a given genus, in a correlator such as Eq.(\ref{scorr}), we will 
have all possible edges (if $J_i$ are enough not to provide a constraint
on the number of edges at a given vertex)\footnote{ We have implicitly
assumed that the correlators we are considering are normal ordered
so that there are no self contractions. See also next footnote.}. 
Hence the generic  
skeleton graph is a triangulation of a genus $g$ surface 
with $n$ labelled vertices (of arbitrary valency). But we should 
remember that
there will always be
contributions from exceptional graphs where one or more faces
are not triangular. The maximum number of Schwinger parameters $\t$
are associated 
with the triangulations. The number of edges in this case, (by an 
application of $V-E+F=2-2g$), is given by $3(n-2+2g)$. As mentioned earlier,
this is independent of the $J_i$ and other details of the correlator. 

This also gives the first indication of the emergence of string moduli. 
The number $6g-6+3n$ is exactly the number of real moduli for a genus $g$ 
Riemann surface with $n$ {\it holes}. Separating out the the $n$ moduli
associated with the sizes of the holes gives the number of moduli of the 
surface with $n$ {\it punctures}. 

In fact, we can argue that the moduli space of skeleton graphs (of genus $g$
with $n$ vertices) is identical to the moduli space $\CM_{g,n}\times R_+^n$.
Consider the generic skeleton graph with triangular faces and look at 
its dual (in the graph theoretic sense). The dual graph has vertices 
associated to each face of the skeleton graph and faces associated to 
each original vertex. And there are dual edges, transverse to the 
original ones, which connect the dual vertices. From the properties of
the generic skeleton graph we can conclude that its dual graph will have 
$n$ (labelled) faces, $6g-6+3n$ edges and cubic vertices. The trivalent
vertices of the dual follow from the triangular faces of the 
(generic) skeleton graph.  
Moreover, we will associate a length $\s={1\over \t} \in (0,\infty)$ 
(``conductance'' in the electrical analogy) 
to each dual edge. Note that the dual graph has the same genus $g$ as 
the original one (since the number of vertices and faces have simply been
interchanged). 

The various inequivalent triangulations are mapped
to inequivalent trivalent graphs. Therefore, associated to each 
skeleton graph of genus $g$ with $n$ vertices, with its set of $\{\t\}$'s,
is a trivalent graph of genus $g$
with $n$ faces and a set of lengths $\{\s\}$ for the dual edges. 
As one sums over inequivalent skeleton graphs, one goes over inequivalent 
trivalent graphs. In fact, we can now better appreciate the role of 
the non-generic skeleton graphs, with 
four-sided (or more) faces. They 
map onto dual graphs with quartic (or higher)
vertices. Such graphs can be thought of as arising when two (or more) cubic 
vertices coalesce, i.e. when the length $\s$ of the edge joining two cubic 
vertices goes to zero. 
In fact, one can continuously move from one trivalent graph
to another inequivalent one
by shrinking some individual edge (``s-channel'') to zero size  and then 
expanding the resultant 
quartic vertex in the other direction (``t-channel''). By this process,
(known to mathematicians as Whitehead collapse) one can connect the 
different inequivalent cubic graphs.

It is a very non-trivial mathematical theorem that the space of 
trivalent fatgraphs of genus $g$ with $n$ labelled faces and a length 
associated to each edge is a cell decomposition of the space
$\CM_{g,n}\times R_+^n$. In other words, as we vary over the lengths
of the edges as well as over the inequivalent graphs we obtain a 
single cover of $\CM_{g,n}\times R_+^n$. Each inequivalent trivalent 
graph fills out a top-dimensional cell in this simplicial decomposition
as we vary the $\s$'s. The graphs with higher point vertices live on 
codimension one, and higher, boundaries of these cells (when one or 
more $\s \rightarrow 0$). At these boundaries the different cells match 
smoothly onto each other.

This theorem is based on the 
work of Mumford, Strebel, Penner and others and may be found in 
Kontsevich \cite{kont}. For a physicist, this statement may be made
plausible by recalling that in cubic open string field theory, the string 
diagrams are made of strips of varying length meeting at cubic
vertices \cite{witten}. 
In fact, it was argued by Giddings, Martinec and Witten \cite{gmw}, 
and later Zwiebach \cite{zwie}, that these diagrams give a single
cover (in our case) of the moduli space, $\widetilde{\CM}_{g,n}$
with $n$ {\it holes}. The mathematical theorem quoted above indicates that 
actually this factors into an $R_+^n$ for 
the diameters of the holes together with the space
of $n$ punctures, $\CM_{g,n}$. 
 
Thus we have argued that
as we vary over the moduli space of skeleton graphs, we 
are covering the appropriate moduli space of string worldsheets.
We should remark here that the process of going to the dual graph 
is in a sense a reflection of open-closed string duality.
In going to the dual graph we are closing off the holes of the original graph
(in replacing them by dual vertices)
while opening up holes/punctures at the original vertices. 
We thus seem to be implementing open-closed string 
duality at least at the level of the geometry of worldsheets\footnote{We 
should point out here that though 
we have been considering skeleton
graphs with triangular faces and their duals, we also need to 
consider skeleton graphs with self-contractions (which are not 
homotopically trivial) at vertices. Because the dual graphs of the latter  
also appear amongst the cells of moduli space. This suggests that 
the prescription of normal ordering which drops such 
self-contractions is perhaps not the natural one from the dual string
point of view.}. In fact, the graph duality and the assignment 
$\s={1\over \t}$ is also natural from the point of view of the 
UV-IR connection. The UV region of the field theory $\t\rightarrow 
0$ is mapped to $\s\rightarrow \infty$ which can be seen to 
correspond to long distance (IR) propagation of the dual closed strings.  
This can also be seen from the expressions for the three point correlators
in Sec. 5 \cite{fads1}.

So we now have a way to understand how free field theory diagrams 
reassemble themselves into closed string worldsheets. It allows us to 
view the expression Eq.(\ref{schem}) we obtained from field theory,
as an integral over the string moduli space $\CM_{g,n}\times R_+^n$.
We also note that this reorganisation of Feynman diagrams can be performed
for the $n$-point function of arbitrary gauge-invariant operators. 
The Schwinger parametrisation and the 
gluing up into skeleton diagrams is something which can be 
always carried out. The general expressions will be more cumbersome
(expressions for general
Schwinger parametrised amplitudes are available in the literature
\cite{lam}) but for specific correlators we can always work them out
explicitly. 
We should add that this reorganisation of field theory 
diagrams can be done for free field theory in any number of dimensions
and with arbitrary matter content (thus not necessarily supersymmetric). 
However, we expect that the interacting
theories will probably have dual string descriptions only in $d\leq4$.

For an interacting theory much of our argument still goes through.
After all the Schwinger 
parametrisation of amplitudes can still be carried out
in the perturbative expansion in 'tHooft coupling $\l$, as also
the simplification into skeleton graphs.
The only difference is that we 
have additional ``internal'' vertices corresponding to the 
interactions. It suggests the appearance of the moduli space 
with additional punctures corresponding to the interactions. Presumably, 
the additional vertex operators associated to the interactions 
then exponentiate (when we sum over the perturbative expansion)
and modify the background. So there is promise 
of extending this approach to the interacting case as well. 

\section{The Integrand on Moduli Space}

Having reorganised the diagrams into a sum over worldsheets, we can take a
closer look at the integrand. Since $\s={1\over \t}$ is the more 
natural variable to describe the cells in moduli space we can rewrite
Eq.(\ref{schem}) (dropping the {\it skel} subscripts) as   
\be
G^{\{J_i\}}_g(k_1, k_2,\ldots k_n)
=\sum_{skel. graphs}\int_0^{\infty}\prod_r d\s_r
{\hat{f}^{\{J_i\}}(\s)\exp{\{-\hat{P}(\s, k)\}}\over\hat{\D}(\s)^{d\over 2}}.
\label{expl}
\ee
where 
\be
\hat{\D}(\s)\equiv\sum_{T_1}(\prod\s)
=(\prod_r\s_r)\D(\t=1/\s)
\ee
and 
\be
\hat{P}(\s,k)\equiv{1\over\hat{\D}(\s)}
\sum_{T_2}(\prod\s)(\sum k)^2=P(\t=1/\s,k)
\ee
are defined in terms of the 1-trees and 2-trees of the skeleton graph
as before but the product in both these 
definitions is over the lines that are {\it not} cut. 
Thus we have an explicit expression, in each cell of the moduli space,
of the integrand. The universal functions $\hat{\D}(\s)$ and $ \hat{P}(\s,k)$
smoothly go from one cell to another at the common cell boundary. 
The multiplicative factor $\hat{f}^{\{J_i\}}(\s)$ which contains, as before,
all the information about the specific operators is, however, 
more sensitive to the 
constraints imposed by the $J_i$'s in the original Feynman graph.

The nett result is that we can write the field theory correlator in the form
mentioned in the introduction, namely, as
\be
G^{\{J_i\}}_g(k_1, k_2,\ldots k_n)
=\int_{\CM_{g,n}\times R_+^n}
[d\s]\rh^{(J_i)}(\s)e^{-\sum_{i,j=1}^nk_i\cdot k_jg_{ij}(\s)}.
\label{finform}
\ee
We can write down $\rh^{(J_i)}(\s)$ and $g_{ij}(\s)$ explicitly in each cell
of the moduli space\footnote{The contributions 
from the non-generic graphs are finite and have support on the boundaries
of the cells. This is reminiscent of similar contributions in open 
superstrings. We thank Ashoke Sen for remarking on this similarity.}. 
We note again the form of the integrand
which is very reminiscent of string theory expressions in flat space. 
Given that we seem to be implementing open-closed string duality,
it is very natural to take the integrand seriously as a candidate 
for correlators in the unknown dual string theory on $AdS$. We will present,
in the next section, some checks in this direction. 

\section{$AdS$ Correlators}

\subsection{Three Point Functions}

For the planar three point function
\be
G^{\{J_i\}}_{g=0}(k_1, k_2, k_3)=\la \Tr\Phi^{J_1}(k_1)
\Tr\Phi^{J_2}(k_2)\Tr\Phi^{J_3}(k_3)\ra_{g=0}
\ee
we can carry out the procedure outlined in the preceding sections. Namely,
we first glue together the multiple lines joining each pair of vertices to
get a skeleton graph which is (for generic $J_i$)
a simple triangle. For this skeleton 
graph we can write an integral over the three effective Schwinger parameters,
as in Eq.(\ref{skel}). In terms of the variables $\s={1\over \t}$, the 
final expression in the case of the three point function is
\be
G^{\{J_i\}}_{g=0}(k_1, k_2, k_3)=\int_0^{\infty}\prod_{r=1}^3
d\s_r\s_r^{(m_r-1)({d\over 2}-1)+{d\over 2}-2}{1\over\hat{\D}(\s)^{d\over 2}} 
\exp{\{-\hat{P}(\s, k)\}},
\label{thrp}
\ee
where in terms of the parameters $\s_i$ for the three edges, we have
\be
\hat{\D}(\s)=\s_1\s_2+\s_2\s_3+\s_3\s_1
\ee
and 
\be
\hat{P}(\s, k)={1\over\hat{\D}(\s)}[\s_1k_1^2+\s_2k_2^2+\s_3k_3^2].
\ee
The multiplicites $m_i$ in Eq.(\ref{thrp}) are determined by the $J_i$ to be:
$m_i={1\over 2}\sum_{k=1}^3J_k-J_i$. 
In this case, since $\CM_{0,3}$ is trivial, the integral in Eq.(\ref{thrp})
is only over the $R_+^3$ factor. 
We can make a change of variables on these three variables to make the 
connection to $AdS$ clear.
\be
{1\over \rh_i}={\s_i \over \hat{\D}(\s)} 
\Rightarrow \s_i={\rh_1\rh_2\rh_3\over 
(\sum_k\rh_k )\rh_i}.
\ee
This change of variables is motivated by the star-delta transformation 
of electrical 
networks. Namely, if $\s_i$ are the conductances 
of a delta or triangle network, 
such as the one we have, then $\rh_i$ are the conductances of the 
equivalent three 
pronged tree or star network. In other words, the $\rh_i$ are the variables 
naturally parametrising the legs of the 
tree one obtains when one glues up the skeleton triangle 
graph. 

Working out the details of the jacobian for this change of variables 
and simplifying the integrand one finds the simple form
\be
G^{\{J_i\}}_{g=0}(k_1, k_2, k_3)=
\int_0^{\infty}\prod_{i=1}^3
d\rh_i\rh_i^{\D_i-{d \over 2}-1}{1\over 
(\sum_k\rh_k )^{\Sigma_k {\D_k\over 2} -{d \over 2}}}
\times e^{-[\sum_{i=1}^3{k_i^2\over \rh_i}]}.
\label{rhthrp}
\ee

We can write this equivalently as 
\be
G^{\{J_i\}}_{g=0}(k_1, k_2, k_3)=
\int_0^{\infty}{dt\over t^{{d\over 2}+1}}
\int_0^{\infty}\prod_{i=1}^3
d\rh_i\rh_i^{\D_i-{d \over 2}-1}t^{\D_i\over 2}e^{-t\rh_i}
e^{-{k_i^2\over \rh_i}}
\label{adsthrp}
\ee
by exponentiating the denominator in Eq.(\ref{rhthrp}).
In position space, taking into account momentum conserving delta functions, 
this becomes
\begin{eqnarray}
G^{\{J_i\}}_{g=0}(x_1, x_2, x_3)&=&
\int_0^{\infty}{dt\over t^{{d\over 2}+1}}\int d^dz
\int_0^{\infty}\prod_{i=1}^3d\rh_i\rh_i^{\D_i-1}t^{\D_i\over 2}
e^{-\rh_i(t+(x_i-z)^2)} \\
&=&\int_0^{\infty}{dt\over t^{{d\over 2}+1}}\int d^dz
\prod_{i=1}^3 K_{\D_i}(x_i,z;t),
\label{adspos}
\end{eqnarray}
where 
\be
K_{\D}(x,z;t)={t^{\D\over 2}\over [t+(x-z)^2]^{\D}}
\label{bulbo}
\ee
is the usual position space bulk to boundary propagator for a scalar field 
in $AdS$, corresponding to an operator of dimension $\D$.
The only slight difference
is that we have parametrised the $AdS$ radial coordinate by $z_0^2=t$.
We see from Eq.(\ref{adspos})
that the $\rh_i$ are indeed parameters for the external legs
of the $AdS$ tree diagram. So the $R_+^3$ integral in the field theory 
has a natural counterpart on the $AdS$ side. 
We can see a similar thing 
for the two point function as well. 

It is encouraging, in trying to answer the second of our questions, 
that the {\it integrands} in these natural parametric 
representations match so well. 
Conformal invariance fixes
the overall functional dependence on the positions. But
this agreement at the level of integrands, {\it in conjunction}
with our general arguments for $n$-point functions, indicates that
we maybe on the right track.
  
The fact that the scalar three point function 
in our procedure could be written purely in terms 
of supergravity bulk-to-boundary propagators is probably
special to this correlation function, especially since
we expect the dual theory to be highly curved.   
We can give a heuristic argument as to why the
full string correlator might simplify in this case. 

Following \cite{pol}\cite{tse} the vertex operator computation,
in an $AdS$ background,
for an $n$-point function of
these scalars would take the form
\be
G^{\{J_i\}}_g(x_1 \ldots x_n)=\int_{\CM_{g,n}}\la\prod_{i=1}^n
K_{\D_i}(x_i,X(\xi_i);t(\xi_i))\ra_{WS}.
\label{adsvert}
\ee
In other words, the vertex operators $\CV(\xi)$ are essentially
the external wave functions of the particles in 
$AdS$ promoted to worldsheet operators. 
Thus $X(\xi), t(\xi)$ denote worldsheet fields for the $AdS$ target space. 
The averaging, as the subscript indicates, is over the worldsheet action for 
these and other fields (including ghosts which would generally also 
enter into the vertex operator).
Using Eq.(\ref{bulbo}) we can rewrite Eq.(\ref{adsvert})
introducing parameters for the external legs as in the case of the three
point function
\be
G^{\{J_i\}}_g(x_1 \ldots x_n)=\int_{\CM_{g,n}}
\int_0^{\infty}\prod_{i=1}^n d\rh_i\rh_i^{\D_i-1}
\la t(\xi_i)^{\D_i\over 2}e^{-t(\xi_i)\rh_i-\rh_i(x_i-X(\xi_i))^2}\ra_{WS}.
\ee
In momentum space, this reads as 
\be
G^{\{J_i\}}_g(k_1 \ldots k_n)=\int_{\CM_{g,n}}
\int_0^{\infty}\prod_{i=1}^n d\rh_i\rh_i^{\D_i-{d\over 2}-1}
e^{-{k_i^2\over \rh_i}}
\la t(\xi_i)^{\D_i\over 2}e^{-t(\xi_i)\rh_i}e^{i k_i\cdot X(\xi_i)}\ra_{WS}.
\label{adsmom}
\ee

In the special case of the $g=0$ three point function, we can 
argue that because of the worldsheet conformal invariance the positions 
of the three vertex operators are irrelevant. The ghost contribution
cancels out the contribution from the non-zero modes of $X(\xi), t(\xi)$, 
so to say. Effectively, only the zero modes contribute and so 
we can replace the worldsheet averaging by an integral over
the zero modes of $X, t$. This is easy to do. The zero mode 
for $X$ just gives the overall momentum 
conserving delta function. That for $t$ is then identical to the 
expression in Eq.(\ref{adsthrp}). In fact, doing the $t$ integral 
goes back exactly to the expression
in Eq.(\ref{rhthrp}) which we had obtained from 
the field theory Schwinger parametrisation after an appropriate change 
of variables on the moduli. 
In other words, carrying out the worldsheet averaging in Eq.(\ref{adsmom})
(for $n=3$) gives Eq.(\ref{rhthrp}).

\subsection{Higher Point Correlators}

The form of the $n$-point function Eq.(\ref{adsmom})
suggests a comparison with Eq.(\ref{finform}) obtained
from our Schwinger parameter procedure. Encouraged by the 
explicit example of the 
three point function, we could try and identify the integral
over the $\rh_i$ in Eq.(\ref{adsmom}) with the 
$R_+^n$ integral in Eq.(\ref{finform}). In fact, the interpretation
of the $R_+^n$ factor as the diameter of the holes also suggests
an identification with the external leg parameters $\rh_i$. 
If this identification is correct, we would be directly 
obtaining the answers for the integrand of Eq.(\ref{adsmom})
from our field theory procedure. 

In any case, a strong check of our conjecture is that our
procedure should give for the integrand on moduli space
an expression which is consistent with all the properties
of a correlator of local operators in some 
two dimensional quantum field theory. This is a strong constraint
since we know that local operators in a field theory obey an OPE.
Various miraculous channel dualities (in spacetime) of string theory  
follow from this OPE. But on the other hand we know that the correlators
in field theory reflect these channel dualities due to the
{\it spacetime} OPE. Since the spacetime OPE is reflected in the Schwinger
parametrised representation, we would like to see it translate into a 
worldsheet OPE. There is some indication that this is the case because
the region of Schwinger parameter space which seems to contribute to 
terms in the spacetime OPE also seems to the region of string moduli space
(via our mapping of the two) where vertex operators come together and one
expects to see a worldsheet OPE. We hope to report on this in the near 
future. 

\section{Conclusions}

So, is free field theory, in general, a string theory?
The universal reorganisation of Feynman diagrams certainly 
gives a strong indication to that effect.
But we will need to study the 
properties of the integrand on moduli space, 
we have obtained, better before we
can give an affirmative answer. As mentioned above, the key point is to
establish a worldsheet OPE.

We would also like to be 
able to extract useful information from this procedure. Perhaps even
reconstruct the worldsheet action. This would be particularly important
if we are to extend the procedure to the perturbative
expansion in the 'tHooft coupling. We would like to see the 
spacetime perturbation theory reassemble itself into a worldsheet 
perturbation expansion which has the effect of changing the background. 

The fact that it is the cubic open string field theory
decomposition of moduli space that appears in our procedure, is perhaps a 
useful hint in understanding the general gauge-string correspondence. 
In other examples where open-closed string duality 
is explicitly realised, open string field theory has often 
made an appearance\cite{gv,gr}.  
Perhaps what we are seeing here is a reflection of that.    

These are some of the many questions thrown up by this approach, 
on which future work will hopefully shed light.




\section*{Acknowledgements}
I would like to thank the organizers of Strings 2004 for 
a very nice conference and for inviting me to present
this work. I must also thank the various participants
for useful discussions.

\end{document}